\documentclass[12pt,a4paper]{iopart}
\usepackage{amssymb,graphicx,cite}
\begin{document}
 
%LATEX FILE OF MANUSCRIPT
%%%%%%%%%%%%%%%%%%%%%%%%%%%%%%%%%%%%%%%%%%%%%%%%%%%%%%%%%%%%%%%%%%%%

%LATEX file of the manuscript
 
%\documentclass[aps,showpacs,twocolumn,floatfix]{revtex4} 
%\documentclass[eqsecnum,aps,twocolumn,epsf]{revtex4} % PH. REV.
%\usepackage{graphicx}
%\documentclass[eqsecnum,aps,twocolumn]{revtex4} % PH. REV.

%\documentstyle[aps,epsf]{revtex}
%\documentstyle[eqsecnum,aps,epsf]{revtex}
%%% <<< epsf commands in the next two lines >>>
%\newcommand{\postscript}[2] {\setlength{\epsfxsize}{#2\hsize}
%\centerline{\epsfbox{#1}}}

%\documentstyle[eqsecnum,aps,epsf]{revtex} % PH. REV. FINAL FORMAT STYLE
%\documentstyle[aps,epsf]{revtex} % PH. REV. FINAL FORMAT STYLE
%%% <<< epsf commands in the next two lines >>>
%\newcommand{\postscript}[2] {\setlength{\epsfxsize}{#2\hsize}
%\centerline{\epsfbox{#1}}}
 
%\documentstyle[preprint,aps]{revtex}
%\documentstyle[eqsecnum,aps]{revtex}
%\documentstyle[aps]{revtex}
%\renewcommand{\baselinestretch}{1.5}
 
%\begin{document}
   
%\twocolumn[\hsize\textwidth\columnwidth\hsize\csname@twocolumnfalse\endcsname
 
\title[Critical number of atoms in an attractive condensate]{Critical
number  
of atoms in 
an attractive Bose-Einstein condensate on an optical plus harmonic traps}
 
\author{Sadhan K. Adhikari}
\address{Instituto de F\'{\i}sica Te\'orica, Universidade Estadual
Paulista, 01.405-900 S\~ao Paulo, S\~ao Paulo, Brazil}

\date{\today}
%\maketitle

\begin{abstract}

The stability of an attractive Bose-Einstein condensate on a joint
one-dimensional optical lattice and an axially-symmetric  harmonic trap 
is studied using the numerical solution of the time-dependent mean-field
Gross-Pitaevskii equation and the critical number of atoms for
a stable condensate is calculated. We also calculate this critical number
of atoms in a double-well potential which is always greater than that in
an axially-symmetric  harmonic trap. 
The critical number of atoms in an
optical trap can be made smaller or larger than the corresponding number
in the absence of the optical trap  by moving a node of the  optical
lattice potential along the axial direction of the harmonic trap. 
This variation of the critical number of atoms can be observed
experimentally and compared with the present calculation.

\end{abstract}

\maketitle

The successful detection \cite{1ex,1rep} of a Bose-Einstein condensate
(BEC) of dilute
weakly-interacting trapped bosonic atoms at ultra-low temperatures
initiated intense theoretical activities on different aspects of the
condensate. The experimental magnetic trap is usually axially symmetric.
More recently in different experiments a periodic optical-lattice
potential generated by a standing-wave laser field has been employed along
the axial direction of the magnetic trap
\cite{1,greiner,cata,ari,cata2,sol}.  The 
optical-lattice potential has been used in the  study of matter-wave
interference
\cite{1}, of oscillating atomic current in a one
dimensional array of Josephson junctions \cite{cata}, of Bloch oscillation
and Landau Zenner tunneling \cite{ari} and 
of
superfluid-insulator
classical \cite{cata2}  and quantum  \cite{greiner} phase transitions 
among
others.
The periodic
optical-lattice potential has also been useful in the generation of
matter-wave
bright soliton \cite{sol}. There are many theoretical studies on a BEC in
a periodic optical-lattice potential \cite{th}.

Most of the above experiments were performed employing  atoms with
repulsive
interaction, where the BEC is stable for any number of atoms. However, 
atoms with attractive
interaction
have
been
used in
some experiments \cite{1rep,sol,exrep,exrep2}, where
the BEC is stable for the number of atoms smaller than a critical number
\cite{crit1,crit2,crit3}.
When the number of atoms increases beyond this critical value, due to
interatomic attraction the radius of the BEC tends to zero and the central
density of the condensate tends to infinity. Consequently, the condensate
collapses while emitting atoms in an exploding fashion due to three-body
recombination until the number of atoms is reduced below the critical
number and a stable configuration is reached \cite{1rep}. With a supply of
atoms from
an external source the condensate can grow again and a series of collapse
and explosion can take place and has been observed in a BEC of $^7$Li
atoms \cite{1rep}.  The attractive condensate has been fundamental in the
generation
of matter-wave bright soliton  \cite{sol} and in the study of collapse and
explosion
of a condensate simulating the supernova explosion of a star \cite{nova}.

The critical number of atoms in an attractive condensate in an
axially-symmetric trap 
has been studied
experimentally \cite{1rep,exrep,exrep2} and 
numerically \cite{crit1,crit2,crit3}
by various authors under different symmetries 
extending from a pancake shaped BEC through spherical to a cigar
shaped one. When the axial trapping frequency $\omega_z$ equals the radial
trapping frequency $\omega_\rho$ one has a spherical BEC. When
$\omega_z>>\omega_\rho$ $(<<\omega_\rho)$ one has a pancake (cigar) shaped
BEC. In view of recent experiments \cite{1,greiner,cata,ari,cata2,sol} on
a cigar-shaped BEC trapped
jointly 
in a
axially-symmetric harmonic plus a optical-lattice trap, we
investigate the stability of an attractive cigar-shaped condensate in such
a trap using the direct numerical solution of the mean-field
Gross-Pitaevskii (GP) equation \cite{8} with appropriate trapping
potential without any
simplifying assumption.  

Specifically, we calculate the critical number of atoms in an attractive
BEC trapped jointly in an axial and optical-lattice traps. This will have
direct consequence on the study of the generation and movement of bright
solitons  as in the recent experiment \cite{sol} with similar traps. We
shall see that this critical number could be increased by appropriately
setting the optical-lattice trap in the axial direction. A larger critical
number is always welcome from the experimental consideration. Also, as the
results of the present mean-field study could be verified experimentally,
this will provide a stringent test of the applicability of the GP equation
on the study of collapse of an attractive condensate in the presence of a
joint optical plus axially-symmetric traps. The most recent
experiment \cite{exrep2} on the critical number in an  axially-symmetric
trap alone 
agrees well with the theoretical
predictions \cite{crit2,crit3}.

There have been many theoretical studies  \cite{dwth} of a BEC in a
double-well 
potential and an experimental study of the same seems to be under control
\cite{dwexp}. In view of this we also present a study of the critical
number of atoms in an attractive condensate in a double-well potential
trap.

The time-dependent BEC wave
function $\Psi({\bf r};\tau)$ at position ${\bf r}$ and time $\tau $
is described by the following  mean-field nonlinear GP equation
\cite{8}
\begin{eqnarray}\label{a} \left[- i\hbar\frac{\partial
}{\partial \tau}
-\frac{\hbar^2\nabla^2   }{2m}
+ V({\bf r})
+ gN|\Psi({\bf
r};\tau)|^2
 \right]\Psi({\bf r};\tau)=0,
\end{eqnarray}
where $m$
is
the mass and  $N$ the number of atoms in the
condensate,
 $g=4\pi \hbar^2 a/m $ the strength of interatomic interaction, with
$a$ the atomic scattering length.  In the presence of the combined
axially-symmetric and optical-lattice traps 
     $  V({\bf
r}) =\frac{1}{2}m \omega ^2(\rho ^2+\nu^2 z^2) +V_{\mbox{aux}}$ where
 $\omega$ is the angular frequency of the harmonic trap 
in the radial direction $\rho$,
$\nu \omega$ that in  the
axial direction $z$, with $\nu$ the aspect ratio, and $V_{\mbox{aux}}$ is
an auxiliar potential simulating a double-well or 
optical-lattice trap introduced later.  
The normalization condition  is
$ \int \rmd {\bf r} |\Psi({\bf r};\tau)|^2 = 1. $

In the axially-symmetric configuration, the wave function
can be written as 
$\Psi({\bf r}, \tau)= \psi(\rho,z,\tau)$.
Now  transforming to
dimensionless variables $x =\sqrt 2 \rho /l$,  $y=\sqrt 2 z/l$,   $t=\tau
\omega, $ with the harmonic oscillator length 
$l\equiv \sqrt {\hbar/(m\omega)}$,
and
${ \varphi(x,y;t)} \equiv   x\sqrt{{l^3}/{\sqrt
8}}\psi(\rho ,z;\tau),$   (\ref{a}) becomes \cite{11}
\begin{eqnarray}\label{d1}
&\biggr[&-i\frac{\partial
}{\partial t} -\frac{\partial^2}{\partial
x^2}+\frac{1}{x}\frac{\partial}{\partial x} -\frac{\partial^2}{\partial
y^2}
+\frac{1}{4}\left(x^2+\nu^2 y^2\right) \nonumber \\
&+&\frac{V_{\mbox{aux}}}{\hbar \omega} -{1\over x^2}  +                                                          
8\sqrt 2 \pi n\left|\frac {\varphi({x,y};t)}{x}\right|^2
 \biggr]\varphi({ x,y};t)=0, 
\end{eqnarray}
where nonlinearity
$ n =   N a /l$. In terms of the 
one-dimensional probability 
 $P(y,t) \equiv 2\pi$  $ \int_0 ^\infty 
\rmd x |\varphi(x,y,t)|^2/x $, the normalization of the wave function 
is given by $\int_{-\infty}^\infty \rmd y P(y,t) = 1.$  

A double-well potential for an axially-symmetric trap can be simulated by
taking 
\begin{equation}\label{dw}
\frac{V_{\mbox{aux}}}{\hbar \omega}\equiv \frac{V_{\mbox{dw}}}{\hbar
\omega} = V_0 \exp (-y^2)
\end{equation} 
where 
$V_0$ is the strength  of the double-well potential $V_{\mbox{dw}} $.  The
optical
potential created with a standing-wave laser field of wavelength
$\lambda$ along the axial direction can be represented by taking  
$ V_{\mbox{aux}}\equiv  V_{\mbox{opt}}=V_0E_R\cos^2 (k_Lz)$,
with $E_R=\hbar^2k_L^2/(2m)$, $k_L=2\pi/\lambda$, and $V_0$  the 
strength \cite{1,greiner,cata,ari}. 
This periodic potential has  a maximum at the center of the
harmonic trap $z=0$. The center of the harmonic trap could also be made to
coincide with a minimum of the periodic potential or anywhere in
between. All these cases can be covered by taking a phase $\delta$ in the
$\cos^2 (k_Lz)$ term so that 
$V_{\mbox{opt}}=V_0E_R\cos^2 (k_Lz+\delta).$  
In terms of the dimensionless laser wave
length $\lambda _0= \sqrt2\lambda/l $ and a  dimensionless 
standing-wave energy parameter $E_R/(\hbar \omega)= 4\pi^2/\lambda _0^2$,
the periodic optical-lattice potential $V_{\mbox{opt}}$ of 
  (\ref{d1}) is
\begin{equation}\label{pot}
\frac{V_{\mbox{aux}}}{\hbar \omega}\equiv 
\frac{ V_{\mbox{opt}}}{\hbar \omega}=V_0\frac{4\pi^2}{\lambda_0^2} 
\left[
\cos^2 \left(
\frac{2\pi}{\lambda_0}y+\delta 
\right)
 \right].
\end{equation}
We shall see that by appropriately choosing  the value of $\delta$ and
$\lambda_0$ the critical number of atoms in an attractive BEC could be
increased after the introduction of the optical trap. 
 
\begin{figure}%[!ht]
 
\begin{center}

\includegraphics[width=.7\linewidth]{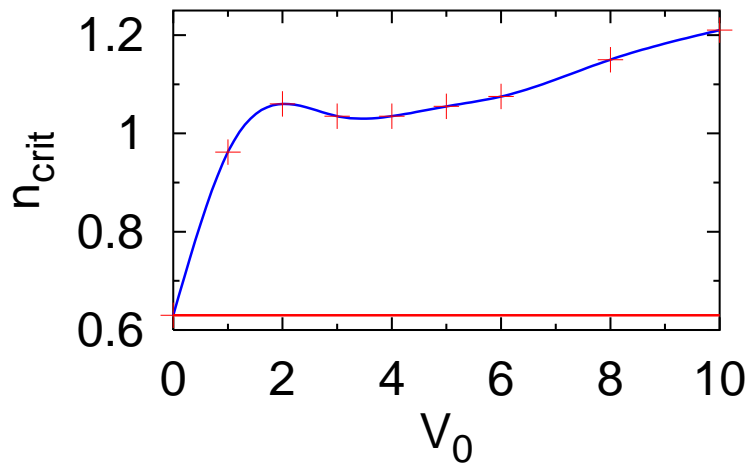}
\end{center}
 
\caption{The reduced critical number $n_{\mbox{crit}}$
  vs. strength $V_0$  of the double-well potential for $\nu =0.5$: 
line with $+ \to$ double well; straight line  $n_{\mbox{crit}}=0.63$   $\to$ $V_0=0$. 
} \end{figure}

We solve   (\ref{d1}) numerically  using a   
split-step time-iteration
method
with  the Crank-Nicholson discretization scheme described recently
\cite{11}.  
The time iteration is started with the  harmonic oscillator solution
of   (\ref{d1}) with
 $n=0$: $\varphi(x,y) = [\nu
/(8\pi^3)  ]^{1/4}$
$xe^{-(x^2+\nu y ^2)/4}$ 
\cite{11}. 
The
nonlinearity $n$  and  the optical-lattice potential parameter $V_0$ 
are  slowly changed  by equal amounts in $5000n$ steps of 
time iteration until the desired value of $n$ and  $V_0$        are
attained for the double-well or the optical-lattice potential. Then,
without changing any
parameter, the solution so obtained is iterated 5000 times so that a
stable
solution  is obtained 
independent of the initial input
and time and space steps. 

For a repulsive condensate   the scattering length $a$ is positive. 
For an attractive condensate the scattering length $a$ and the
nonlinearity $n$ are negative.  In the calculation we find that, for
an attractive condensate with negative $n$, stable
solution of the GP equation cannot be obtained for $|n|$ greater 
than a 
critical value $n_{\mbox{crit}}$ $-$ the reduced critical number. For the
spherically symmetric case 
$n_{\mbox{crit}}= 0.5746$ \cite{crit2}.  The critical value for the number
of atoms is
given by $N_{\mbox{crit}}=  n_{\mbox{crit}}l/|a|.$ The dimensionless
parameter  $n_{\mbox{crit}}$ is universal, whereas the number
$N_{\mbox{crit}}$ depends on the typical experimental set up, e.g., the
atomic scattering length and the harmonic oscillator length of the trap.

\begin{figure}%[!ht]
 
\begin{center}

\includegraphics[width=.49\linewidth]{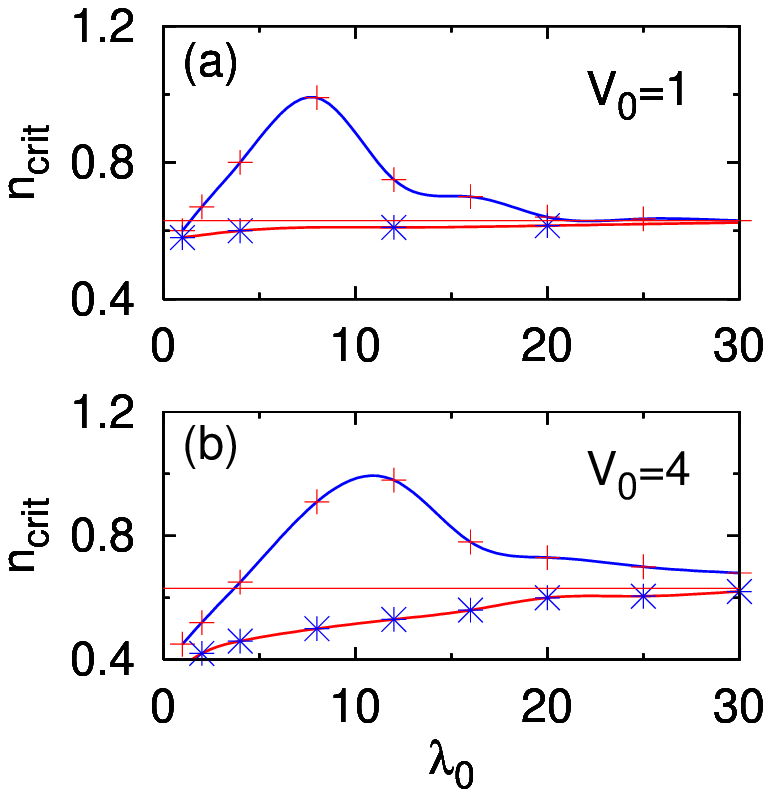}
\includegraphics[width=0.49\linewidth]{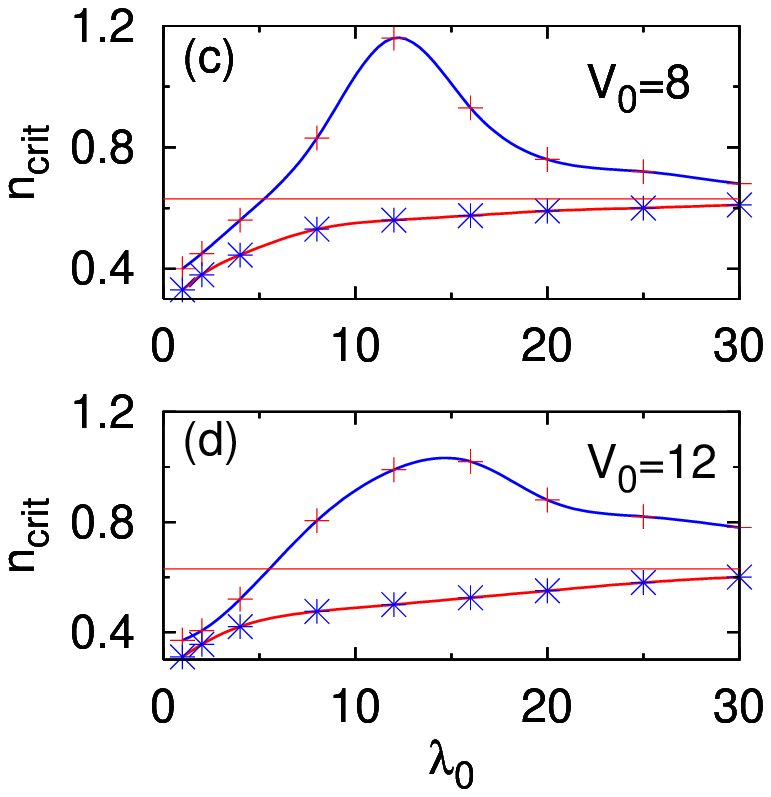}
\end{center}
 
\caption{The reduced critical number $n_{\mbox{crit}}$
  vs. 
dimensionless wave length $\lambda_0$  for $\nu =0.5$
and 
(a) $V_0=1$, (b)
 $V_0=4$, (c)  $V_0=8$, and  (d) $V_0=12$: line with $+ \to$ $\delta
=0$;  line with $\star \to$ $\delta
=\pi/2$; straight line  $n_{\mbox{crit}}=0.63$   $\to$ $V_0=0$. 
} \end{figure}

First we present a study of the critical number  $n_{\mbox{crit}}$     
for the  double-well potential. We consider an elongated cigar-shaped
harmonic trap with $\nu =0.5$  although the calculation can be
extended for other values of $\nu$. In figure 1 we plot  $n_{\mbox{crit}}$
vs. $V_0$.
The case  $V_0=0$ corresponds to the absence of the
optical-potential or double-well  trap and in this case we obtain
$n_{\mbox{crit}}=0.63$ in
agreement with 
 $k_{\mbox{crit}} \equiv 
n_{\mbox{crit}} \nu^{1/6}= 0.56 $ reported in reference
\cite{crit2}
for $\nu =0.5$.  We find from figure 1 that $n_{\mbox{crit}}$ for the
double-well potential is always larger than that for the
corresponding harmonic axial
trap. The double-well trap separates the original condensate
effectively into two parts with a region of low density in between. When
$V_0 $ is large this separation is more pronounced and  one
effectively has two separated condensates and the critical number of atoms
in the double-well trap 
tends approximately to twice the critical number in the corresponding  
harmonic trap alone. For $V_0=0$
the double-well is absent and  $n_{\mbox{crit}}=0.63$. Figure 1 shows the
evolution of  $n_{\mbox{crit}}$
between these two limiting values. It should be recalled that the exact 
limiting value for  $n_{\mbox{crit}}$   for large $V_0$ as well as the
evolution between the two limits will depend on the type of the
double-well
potential employed. Figure 1 represents these general features for the
double-well 
potential 
(\ref{dw}).

In most experimental set ups with an optical-lattice potential a cigar
shaped trap has been used and we consider only  this symmetry
($\nu<1$) in this paper.
Here we consider   the anisotropy parameter  $\nu =0.5$ and  four
values of
the strength $V_0$ of the optical-lattice potential: $V_0=1, 4, 8$ and
12. We consider the two limiting cases $\delta =0$ and $\pi/2$ in the
optical-lattice periodic potential (\ref{pot}).

  The critical nonlinearity $n_{\mbox{crit}}$ vs. dimensionless wave
length $\lambda_0$ for different $V_0$ and $\delta$ $ (=0,\pi/2) $ are
plotted in figures 2. 
This constant critical value ($n_{\mbox{crit}}=0.63$) for $\nu=0.5$ and
$V_0=0$ is
also shown in figures 2
for a comparison. The lines with + represent $\delta=0$ or the
$\cos^2(2\pi y/\lambda_0)$ dependence of $V_{\mbox{opt}}$ whereas the
lines with $\star$  represent $\delta=\pi/2$  or the
$\sin^2(2\pi y/\lambda_0)$ dependence.  For a fixed $V_0$ and $\delta=0$
one has   $n_{\mbox{crit}}<0.63$ for small $\lambda_0$. With the increase
of $\lambda_0$,  $n_{\mbox{crit}}$ increases past 0.63 and eventually it
decreases to 0.63 asymptotically for a large enough $\lambda_0$. For
$\delta=\pi/2$ and small  $\lambda_0$,   $n_{\mbox{crit}}$ starts at a
value 
very similar to the  $\delta=0$   case. With the increase of $\lambda_0,$
$n_{\mbox{crit}}$ monotonically increases in this case to the value 0.63,
asymptotically, however,  remaining always smaller than 0.63. The plots of
$n_{\mbox{crit}}$ vs. $\lambda_0$ 
for any $\delta$ lie between the $\delta=0$ and $\delta=\pi/2$ limits
shown in figures 2.

\begin{figure}%[!ht]
 
\begin{center}

\includegraphics[width=0.49\linewidth]{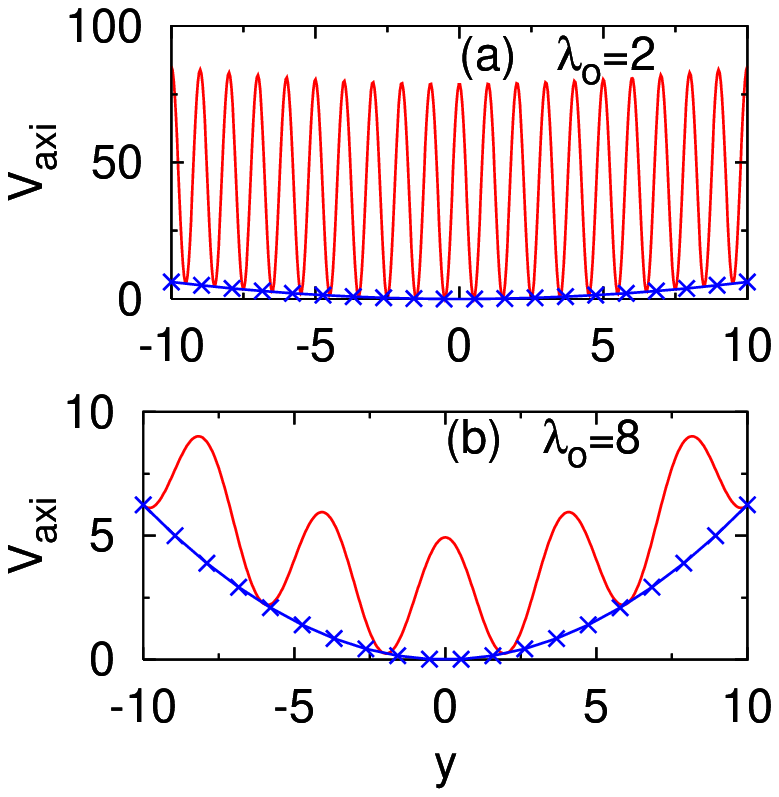}
\includegraphics[width=0.49\linewidth]{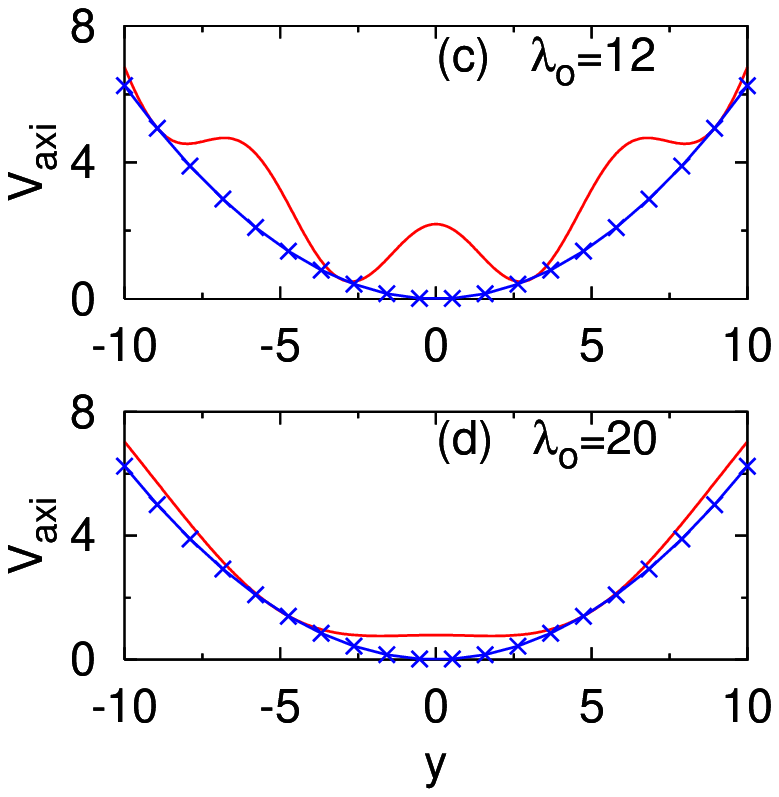}
\end{center}
 
\caption{The full axial potential (\ref{tx}) 
$V_{\mbox{axi}}$ vs. $y$ for $V_0=8$, $\delta = 0 $ and $\nu=0.5$ for 
(a) $\lambda_0=2$,
(b) $\lambda_0=8$,
(c) $\lambda_0=12$, and 
(d) $\lambda_0=20$: full line $\to$ $V_{\mbox{axi}}$;
full line with $\times$ $\to$ axial harmonic potential 
$\nu ^2 y^2/4$.
} \end{figure}

\begin{figure}[!ht]
 
\begin{center}

\includegraphics[width=0.49\linewidth]{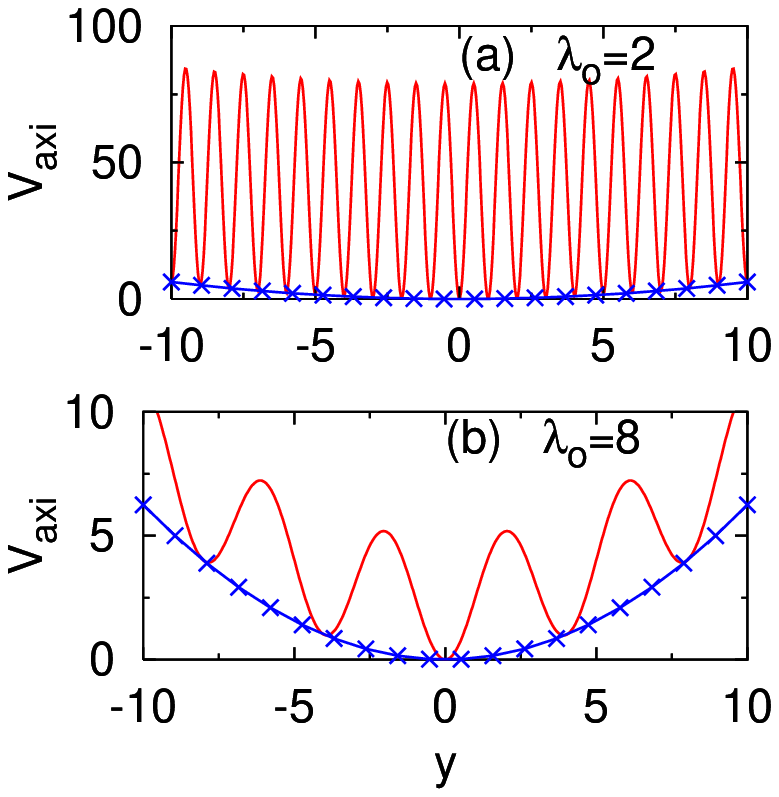}
\includegraphics[width=0.49\linewidth]{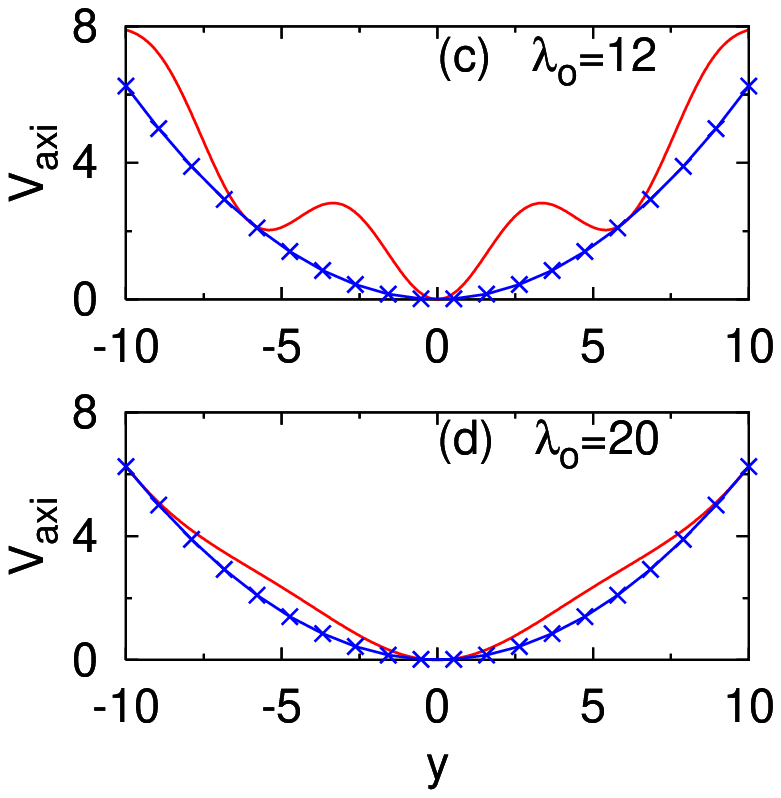}
\end{center}
 
\caption{Same as in figures 3 but for $\delta =\pi/2$
} \end{figure}

The behavior of the  $n_{\mbox{crit}}$ vs.  $\lambda_0$ curves reported in
figures 2 can be understood from a consideration of the total axial
potential 
\begin{equation}\label{tx}
V_{\mbox{axi}}= \frac{1}{4}\nu^2 y^2+ 
V_0\frac{4\pi^2}{\lambda_0^2} 
\left[
\cos^2 \left(
\frac{2\pi}{\lambda_0}y+\delta 
\right)
 \right].
\end{equation}
It should be noted that in the presence of the periodic optical-lattice
potential the radial trap remains unchanged as the optical lattice does
not provide any confinement in the radial direction.  In figures 3 we plot
$V_{\mbox{axi}} $ vs. $y$
for four different values of wavelength $\lambda_0$ for $V_0=8$ and
$\delta=0$.  
For $\lambda_0=2$ 
we have a large number of nodes in the potential $V_{\mbox{axi}}$ 
revealing a large number
of confining traps of the periodic potential as one can find in
figure 3 (a). The number of 
available  traps of $V_{\mbox{axi}}$ in a certain region of space in the
axial $y$ direction
decreases as $\lambda_0$ increases (to 12 through 8) as found in figures 3
(b)  and (c). Eventually, for a large enough
$\lambda_0$ ($\sim 20$) the  confining traps of the periodic potential
$V_{\mbox{axi}}$
disappear and it reduces 
essentially to the  harmonic trap potential as in
figure 3 (d). For large
$\lambda_0$ the extention of each of the periodic traps of
$V_{\mbox{axi}}$
is larger than that of the axial harmonic trap, and $V_{\mbox{axi}}$
tends towards the  axial harmonic trap $\nu ^2 y^2/4$.
Hence one can
safely conclude that
for large  $\lambda_0$, $n_{\mbox{crit}}$ tends to its value $(=0.63)$
 in the
absence
of the optical potential as one can find from figures 2. 

The typical axial
dimension of the attractive condensate before collapse is larger than the
width of a
single confining trap of the optical potential for small $\lambda_0$
(figure 3 (a)), hence the condensate should occupy a few of these
confining optical-lattice traps. Consequently, the BEC will be squezeed
towards the center 
of each these confining traps with a region of low density between the
traps.  Because of this further squeezing the BEC will collapse with a
smaller number of atoms and lead to a $n_{\mbox{crit}}$ smaller than 
that in the absence of the optical trap.   This effect will be large for a 
stronger optical potential with large $V_0$ as one can see in figures 2.
The typical dimension of the attractive condensate in the axial direction 
is smaller or comparable to the dimension of a single trap of the optical
potential for larger $\lambda _0$. 
For $\delta=0$ the appearance of a maximum of the axial potential
 $V_{\mbox{axi}}$ 
at  $y=0$ pushes the condensate apart into two nearby minima of the axial
potential. Consequently, in these cases the BEC swells and  one
effectively has two
separated pieces of
the condensate
in two traps. 
In this case the central part of the total trap, where the
BEC is formed,  is similar
to a double-well potential.
Consequently,  there is more space available for the full 
condensate and hence
$n_{\mbox{crit}}$ can grow beyond the critical value  in the absence of
the optical trap.

In figures 4 we plot $V_{\mbox{axi}} $ vs. $y$ for four
different values of wavelength $\lambda_0$ for $V_0=8$ and
$\delta=\pi / 2$.  In this case  one has a minimum of
the axial potential   $V_{\mbox{axi}}$
at  $y=0$ in contrast to  the $\delta=0$ case where one has a maximum of
$V_{\mbox{axi}}$
at  $y=0$.
Consequently, for
larger $\lambda _0$ the BEC will be   essentially  squeezed into
the central
trap
of the optical-lattice potential. The central part of the total potential,
where the BEC is formed, 
is never of the double-well type, but rather like a stronger single-well
type.  Because of the above  squeezing the
critical number of
atoms will  always be less than the critical number in the absence of the
optical-lattice trap. Again $n_{\mbox{crit}}$  grows
monotonically between the small and large $\lambda_0$ limits as one can
see in figures 2.

In conclusion, using the  numerical
solution
of the  GP equation 
we have studied the stability of an attractive BEC trapped jointly by an
optical and an axially-symmetric harmonic traps. We calculate the critical
number of atoms for stability in this case. We find that the reduced
critical number $n_{\mbox{crit}} = N_{\mbox{crit}} |a|/l$  could either
increase or decrease in relation to the same number in the absence of the
optical trap. However, for most experimental set ups $\lambda_0$ is
expected to be less than 3 and and from figures 2 we find that  
$n_{\mbox{crit}}$ should reduce after the
introduction of the optical lattice trap.  For a double-well potential
$n_{\mbox{crit}}$ is always
larger than the same for the corresponding   axially-symmetric  harmonic
trap. 
A prior knowledge of  the critical number $n_{\mbox{crit}} $ 
as obtained in this paper will be useful in the experimental generation
and study of matter-wave bright solitons under similar trapping
conditions. Also,  $n_{\mbox{crit}} $ can be measured experimentally and
compared with the present theoretical results. This will be helpful in
evaluating the applicability of the mean-field GP equation in the study of
attractive condensates trapped in a joint periodic  optical plus a
harmonic potential.

\vskip 0.5cm

\ack  

The work was supported in part by the CNPq and FAPESP
of Brazil.

 \end{document}